\documentstyle[multicol,pra,aps,amsfonts,epsf]{revtex}

\newcommand{\beq}{\begin{equation}}
\newcommand{\eeq}{\end{equation}}
\newcommand{\beqy}{\begin{eqnarray}}
\newcommand{\eeqy}{\end{eqnarray}}

\newenvironment{Definition*}{{\bf Definition}}{}

\makeatletter
\def\@beginTheorem#1#2{\trivlist \item[\hskip \labelsep{\bf #1\ #2}]}
\def\@opargbegintheorem#1#2#3{ \trivlist
      \item[\hskip \labelsep{\bf #1\ #2\ (#3)}]}
\makeatother
\makeatletter
\def\@beginLemma#1#2{\trivlist \item[\hskip \labelsep{\bf #1\ #2}]}
\def\@opargbeginLemma#1#2#3{ \trivlist
   \item[\hski
 Hence we have the same statements
about the increase of the supports for increasing depth, where
the local transformations are not counted for the depth.
p \labelsep{\bf #1\ #2\ (#3)}]}
\makeatother

\makeatletter
\def\@beginDefinition#1#2{\trivlist \item[\hskip \labelsep{\bf #1\ #2}]}
\def\@opargbeginDefinition#1#2#3{ \trivlist
      \item[\hskip \labelsep{\bf #1\ #2\ (#3)}]}
\makeatother

\makeatletter
\def\@beginCorollary#1#2{\trivlist \item[\hskip \labelsep{\bf #1\ #2}]}
\def\@opargbeginCorollary#1#2#3{ \trivlist
      \item[\hskip \labelsep{\bf #1\ #2\ (#3)}]}
\makeatother

\makeatletter
\def\@beginExample#1#2{\trivlist \item[\hskip \labelsep{\bf #1\ #2}]}
\def\@opargbeginExample#1#2#3{ \trivlist
      \item[\hskip \labelsep{\bf #1\ #2\ (#3)}]}
\makeatother

\def\C{{\mathbb{C}}}

\def\N{{\mathbb{N}}}

\title{Performing joint measurements and transformations on  several  
qubits\\
by operating on a single `control' qubit}

\author{Dominik Janzing\thanks{Electronic address: janzing@ira.uka.de}, 
Thomas Decker, 
   and 
Thomas Beth}
\address{Institut f\"ur Algorithmen und Kognitive Systeme, 
Universit\"at Karlsruhe, Am Fasanengarten 3a,\\
    D--76\,131 Karlsruhe, Germany}

\begin{document}

\maketitle

\begin{abstract}
An $n$-qubit quantum register  can in principle be completely
controlled by operating on a single qubit that interacts
with the register via an appropriate fixed interaction.
We consider a hypothetical system consisting of 
$n$ spin-1/2 nuclei that interact with an electron spin
via a magnetic interaction. We 
describe algorithms that measure non-trivial joint observables
on the register by acting on the control spin only.
For large $n$ this is not an efficient model for
universal quantum computation but it can be modified
to an efficient one if one allows $n$ possible positions
of the control particle.

This toy model of measurements illustrates in which way
specific interactions between the register and a probe 
particle {\it support}
specific types of joint measurements in the sense that some
joint observables can be measured by {\it simple} sequences of operations
on the probe particle.
\end{abstract}

\begin{multicols}{2}

\section{Introduction}

Quantum control theory is an important part of modern
quantum information theory and a basic field
of quantum computation. Actually, every  model of
a quantum computer has to be justified by control theoretic considerations
as its elementary transformations have to be achievable 
by controlling parameters of a real physical system.
Even the definition of quantum complexity 
relies on the available {\it resources} for computation. 
The resources may for instance  be
one- or two-qubit gates \cite{DiV},  bipartite interactions \cite{BCL}
a multi-particle interaction \cite{Martineffizient}, 
certain types of measurements \cite{RB00,LeuMess}  or whatever.
To understand what is easy and what is difficult to compute 
in nature various models have to be studied.

In the control theoretic model studied here we consider 
$n$ qubits that are addressed and controlled by a quantum controller
that is given by a single qubit. We assume that there is no possibility
to directly address the $n$-qubit register.
In Sections \ref{Group},\ref{Simple}, and \ref{Gates} 
we assume that each of the $n$ qubits that should be controlled
interact with the control qubit via a fixed interaction.

The main part of this paper describes procedures
for measuring joint observables. The usual approach of quantum computation
is that single qubit measurements are elementary and that 
measurements of arbitrary joint observables can be reduced to those
``basic'' measurements by implementing an appropriate unitary operation
in advance. More explicitly, this can be done as follows:

If the observable $A$ is non-degenerate, i.e., it has $2^n$ different 
eigenvalues, one can implement a unitary $U$ with the property that 
 $UAU^\dagger$ is diagonal with respect to any basis that consists
only of tensor product states and measure in this basis afterwards. In 
case one wants to achieve that the system is in an eigenstate 
of $A$ afterwards one can implement $U^\dagger$ after the measurement.
For degenerate $A$ with spectral
decomposition $A=\sum_{j\leq k} \lambda_j P_j$
 one can use an ancilla register having at least $k$
basis states and implement a   unitary transformation 
$V$ such that
\[
V(|\psi\rangle \otimes |0\rangle)= 
\sum_j P_j |\psi \rangle \otimes |j\rangle\,,
\]
for all register states $|\psi\rangle$
where $|0\rangle$  is a fixed initial state of the ancilla register and 
$(|j\rangle)_{j\leq k}$ are mutual orthogonal vectors.
Then one can measure the ancilla system and obtains
a measurement of the observable $A$ with corresponding state reduction
in the sense of the von-Neumann 
projection postulate.

This method to measure joint observables reduces the question
of their {\it complexity} completely to the complexity of the required
unitary transformations. In this model the single  qubit
observables are the most elementary ones by definition.
In contrast, we emphasize that every measurement, whether it is a
single-qubit observable or a complex joint observable, relies on interactions
between the considered system with the environment (the 
measurement apparatus). Depending on the structure of the interaction
between register and measurement apparatus the time evolution may
transport {\it directly} some information about the {\it joint} 
state of the register
to the apparatus. The question studied here is how to construct
measurement algorithms for joint observables and we 
show some examples of non-trivial joint observables
that can be measured by comparably simple control operations.
In other words,
we would like to know which kind of information transfer to the measuring 
device is ``done by the interaction'' without much external control.
Therefore we assume to have a fixed interaction between probe particle
and measuring device. 

We do not claim the  model used here to be ``the ultimate setting''
to discuss the complexity of joint measurements.  
However, it has some nice features:

\begin{itemize}
\item The measuring device is as small as possible (with respect to its
register size). If the measuring device would be a large system
too the set of observables that are easy to measure would depend
on the set of basic observables of the measuring device.
To avoid this kind of circularity in treating the question of complexity
of observables we have to keep the measuring device small.

\item The set of extern control possibilities is extremely restricted
as we only allow one-qubit operations on the control spin.
If transformations on the register would be allowed the complexity of
measurement procedures would rather depend on the set of available 
transformations on the register than on the interaction with the environment.

\end{itemize}

Processes where the ``measuring device''
is a quantum mechanical probe particle are sometimes called
{\it premeasurements} as they do not describe the interface between
the classical and the quantum world (we do not address this deep
problem here).
The simplicity of the measuring device makes it possible 
to describe the instructions of the measurement process as
a sequence of operations in the unitary group $SU(2)$.
In particular, the use of 
control operations in finite subgroups of $SU(2)$
makes is possible to formulate the design of
measurement processes as a group theoretical problem.
More explicitly, we 
assume the interaction between the single qubit and the register to be
given by 

\begin{equation}\label{Wwe}
H:=\sum_\alpha \sigma_\alpha \otimes  \sum_j c_j \sigma_\alpha^{(j)}
\end{equation}

where $\sigma_\alpha^{(j)}$ for $\alpha=x,y,z$ is the Pauli matrix 
$\sigma_\alpha$ 
acting on 
register  qubit $j$ and the matrices $\sigma_\alpha$ on the left hand side 
of eq. (\ref{Wwe}) operate on the Hilbert space of the control qubit.
For the
following hypothetical system this Hamiltonian is an appropriate
approximation:
Let the register consist of nuclear spins and the controller be
an electron spin. Due to different gyro-magnetic factors, the
magnetic moment  of a nucleus
is considerably smaller than the magnetic moment of an electron.
Therefore we neglect the magnetic interactions among the nuclear spins.
We assume that the nuclei have different distances $d_j$ from the electron.
Hence they interact with different interaction strength $c_j$ (see 
Fig. 1).

\begin{figure}\label{3Sp}
\epsfbox[-40 0 130 135]{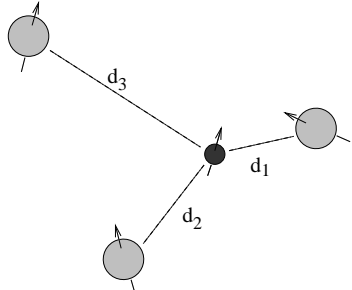}
\caption{Three nuclear spins interacting with an electron spin. The distances
$d_1,d_2,d_3$ determine the interaction strengths $c_1,c_2,c_3$.}
\end{figure}

Abstractly speaking, we have a composed system
with Hilbert space
\[
\C^2\otimes \C^N\,,
\]
where $N:=2^n$,
an interaction Hamiltonian
\begin{equation}\label{Inter} 
H:=\sigma_x \otimes S_x +\sigma_y \otimes S_y +\sigma_z \otimes S_z\,,
\end{equation}
where $S_x,S_y,S_z$ are self-adjoint operators on $\C^{N}$.
Furthermore we assume that 
unitary operations $u\otimes {\bf 1}$ on the controller
can be implemented arbitrarily fast, i.e., the time  required for the 
measuring process is essentially determined by the interacting
between register and probe particle.

In Section \ref{Group} we describe how to measure
the observable $f(S_\alpha)$ if $f$ is an arbitrary two-valued
function on the spectrum of $S_\alpha$ with $\alpha=x,y,z$.
 
In Section \ref{Simple} we present some examples for joint observables that 
are simple to measure in our setting. In Section \ref{Gates}
 we show that usual two qubit gates can be implemented efficiently if
the strength of the interaction between different qubits and the controller
can be varied, for instance when the position of the probe particle 
changes.

\section{Constructing premeasurements using finite groups}

\label{Group}

First we describe two simple tricks to convert the interaction
\[
H:=\sum_\alpha \sigma_\alpha \otimes S_\alpha
\]
to one of the interactions $\sigma_\alpha \otimes S_\alpha$ 
by operating on the controller only: The first method is to
intersperse the natural time evolution
by fast implementations of $i\sigma_\alpha$-rotations on the controller.
If the time intervals are made arbitrarily small, the interaction converges
to the average Hamiltonian $\sigma_\alpha\otimes S_\alpha$
(``selective decoupling'' techniques, e.g. \cite{Leung}).
Similarly, one can apply a strong magnetic field in $x,y,$ or $z$
direction at the position of the electron spins.
To see that a strong field in $z$-direction cancels the
$\sigma_x$ and $\sigma_y$-terms note that
the time average of $\exp(i\sigma_z B t) \sigma_x \exp(-i\sigma_z Bt)$
over the interval $[0,2\pi/B]$  vanishes. Hence the 
interaction term $\sigma_x\otimes S_x$ 
is approximatively cancelled as long as $B \gg \|S_x\|$.
The same holds for $\sigma_y \otimes S_y$.

Now we describe how to measure two-valued observables $f(S_z)$ 
using the interaction $\sigma_z\otimes S_z$ if $f(S_z)$ denotes
the operator obtained by application of $f$ in the sense
of operator functional calculus.
In the following we assume that $S_z$ has the 
eigenvalues $0,1,2,\dots,N-1$. 
Up to a translation and a time scaling factor, this is the case
when $c_j=2^j$. Note  that the translation of the spectrum of $S_z$ 
corresponds to an additional term proportional to 
$\sigma_z\otimes {\bf 1}$ 
for the probe spin Hamiltonian that is not mentioned explicitly 
in the following.
 
The interaction strength $c_j$ may  be achievable by appropriate 
distances
$d_j$ between register spin $j$ and the controller spin.

After the time $t$ the system evolution is given by
\[
\exp(-it \sigma_z \otimes S_z )\,. 
\]
If the register is in an eigenstate with eigenvalue $j$ the controller
evolves according to
\[
\exp(-ij\sigma_z t) \,.
\]
We describe therefore the total time evolution
as 
\begin{eqnarray*}
&&({\bf 1},
\exp(-i\sigma_zt),\exp(-i2\sigma_zt),\dots,\exp(-i(N-1)\sigma_zt)) \\
&&\,\,\,\,\,\,\,\in SU(2)\times SU(2) \times\dots \times SU(2)\,
\end{eqnarray*}
and external unitary operations $u\otimes {\bf 1}$  (where ${\bf 1}$ denotes teh identity) 
on the controller
as
\[
(u,u,\dots,u) \in SU(2)\times SU(2) \times\dots \times SU(2)\,.
\]
In both cases,
the $j$-th component (note that the counting begins with $0$)
of the direct product describes the
 transformation if eigenvalue $j$ is present. 

The evolution caused by the interaction
has therefore the abstract form
\[
(v^0,v^1,v^2,\dots,v^{N-1})\,,
\]
with $v\in SU(2)$. It is a {\it conditional} transformation on
the control qubit depending on the state of the register.

Note that every transformation of this form can be 
implemented: Conjugate the
conditional transformation $\exp(-ijt\sigma_z)$ 
by an unconditional operation $w$ on the controller and obtain
the conditional transformation $w \exp(-ijt\sigma_z) w^\dagger$.

Let $f:\{0,1,2,\dots,N-1\}\rightarrow 
\{0,1\}$ be an arbitrary function. Then
an algorithm for measuring $f(S_z)$ must have the property
that eigenvalues $k,l$ of $S_z$ with $f(k)\neq f(l)$ lead to
mutual orthogonal controller states and 
to equal controller states if $f(k)= f(l)$.
We can achieve this by a sequence of conditional and unconditional
transformations that implements the conditional 
transformation $u^{f(j)}$ where $u$ is an element of order $2$, i.e.,
$u\neq {\bf 1}$ and $u^2={\bf 1}$. We interpret the transformations 
as rotations of the controller's Bloch vector. Then
we describe each step of the measurement procedure as
an element of 
\[
SO(3)\times SO(3)\times \cdots \times SO(3)\,,
\]
and the resulting transformation is
a rotation by $180$ degree in the $j$-th component of the direct product 
if and only if $f(j)=1$.
One can initialize the controller spin in such a way that its Bloch 
vector is orthogonal to the rotation axis and achieves
that the resulting states of the controller are perfectly distinguishable
for different values  $f(j)$.
This justifies the following definition: 

\vspace{0.3cm}
\noindent
{\bf Definition}
Let $f$ be a function $\{0,1,2,\dots,N-1\}\rightarrow \{0,1\}$.
An algorithm for measuring $f$ with $k$ steps is a sequence
\[
W_1,W_2,\dots,W_k
\]
where each $W_j$ is of the form  $(g,g,\dots,g)$ with $g\in SO(3)$ or
$(v^0,v^1,\dots,v^{N-1})$ where $v$ is an arbitrary element of
$SO(3)$ 
such that
\[
W_k W_{k-1} \cdots W_1 = (u^{f(0)},u^{(f(1)},\dots,u^{f(N-1)})
\]
with $u^2={\bf 1}$ and $u\neq {\bf 1}$.

\vspace{0.3cm}
Note that it is irrelevant which element $v$ of order $2$ 
is chosen since they can be transformed into each other by conjugation
with unitaries on the controller. The construction of this kind
of measurement procedures has been discussed in \cite{JZAB} by
Lie-algebraic methods. Here we are interested in a more explicit
design of procedures that does not refer to an infinite number
of infinitesimal transformations.
 
If one restricts the attention to conditional 
and unconditional transformations in a finite group $G$, the construction
of measurement procedures reduces to a word problem in the group
$G\times G\times \cdots \times G$ and computer algebra software
can be used to solve it. We checked, for instance, that
the elements $(g^0,g^1,\dots,g^{29})$ and $(g,g,\dots,g)$ with $g$ in 
the alternating group  $A_5$
generate the whole group $A_5\times A_5 \times \cdots \times A_5$.     
This means that every function on $30$ equidistant eigenvalues
can in principle be measured by 
$A_5$-transformations on the controller.
The group $A_5$ is the symmetry group of dodecahedron and the icosahedron.
If $u\in A_5$ has the order $2$ there exists two opposite 
corners of the dodecahedron or icosahedron that are permuted
by $u$. If the probe spin is initialized to one of them one can 
distinguish the transformations $u$ and ${\bf 1}$ 
by measuring the resulting 
state. 

Now we show that a measurement process exists for every $N$ and every 
two-valued 
function $f$
if one does not restrict the attention to a specific finite group
but combine transformations of different finite groups.
The following argument shows that it is sufficient to
construct measurements for  the functions $f_i$ with
$f_i(j):=\delta_{ij}$, where $\delta_{ij}$ denote the Kronecker symbol.
An arbitrary function $f$ can be written as
\[
f:=\oplus_{i\in I} f_i\,,
\]
where
$I$ is appropriate subset  of  $\{0,1,2,\dots,N-1\}$.
Since the XOR operation $\oplus$ corresponds to a concatenation of
measurement procedures we can restrict our attention to each $f_i$.

We construct  measurements for all $f_i$ by the following recursive scheme:
We assume that we have already constructed 
a procedure for measuring the function $f_{i;k}$ that is defined
by $f_{i;k}(j)= 1$ if and only if $i=j \,(mod \,\,2^k)$. 
Now we describe the measurement for $f_{i;k+1}$.
Let the measurement procedure for $f_{i;k}$ 
implement 
\[
(v^{f_{i;k}(0)},v^{f_{i;k}(1)},\dots,v^{f_{i;k}(N-1)})\,,
\]
where $v$ is an arbitrary 180 degree rotation.

Let $d$ be a rotation by $2\pi/2^{k+2}$, i.e.,
\[
d^{2^{k+2}}={\bf 1}\,.
\]
around an axis orthogonal to the axis of $v$.

Then the recursion runs as follows:

\begin{enumerate}

\item
Implement the conditional transformation
\[
(d^0,d^1,\dots,d^{N-1})\,.
\]

\item 
Implement the unconditional transformation
\[
(d^{-i-2^k},d^{-i-2^k},\dots,d^{-i-2^k})\,.
\]

\item
Call the measurement procedure for $f_{i;k}$, i.e., implement
\[
(v^{f_{i;k}(0)},v^{f_{i;k}(1)},\dots,v^{f_{i;k}(N-1)})\,.
\] 

\item
Implement the inverse of the transformation in 1, i.e.,
\[
(d^{-0},d^{-1},\dots,d^{-N+1})\,.
\]

\item 
Implement the unconditional transformation
\[
(d^{i+2^k},d^{i+2^k},\dots,d^{i+2^k})\,.
\]

\item
Call the measurement procedure for $f_{i;k}$ again.

\end{enumerate}

The resulting transformation implements on the $j$-th component
the $SO(3)$-rotation
\begin{equation}\label{Ausdr}
v^{f_{i;k}(j)} d^{i-j+2^k}  v^{f_{i;k}(j)} d^{j-i-2^k}   \,.
\end{equation}
For every $j\neq i \,(mod \,\, 2^k)$ the rotation is the identity
since expression (\ref{Ausdr}) reduces to
\[
d^{i-j+2^k} d^{j-i-2^k} ={\bf 1}
\]
due to
\[
v^{f_{i;k}(j)}={\bf 1} \,.
\]
Now we distinguish between the two remaining cases
\[
j=i+(2l+1)\, 2^k  \hbox{ and } j=i+ (2l) \,2^{k}\,.
\]

In the first case
expression (\ref{Ausdr}) reduces to
\[
 v \, d^{-2^{k+1}l} \, v \,d^{2^{k+1}l} \,.
\]
This is the identity since 180 degree rotations around 
mutual orthogonal axis commute.

For  $j=i+ (2l)\, 2^{k}$
expression (\ref{Ausdr}) is equal to
\[
v \,d^{2^k-2^{k+1}l} \,v \,d^{2^{k+1}l-2^k}  = v\, d^{2^k} \, v \,d^{-2^k}
\]
since 
$
d^{-2^{k+1}l}
$
commutes with $v$.

Hence we obtain a rotation by 180 degree since 
$
d^{2^k}
$
is a 90 degree rotation and the commutator 
between a 180 degree rotation and
a 90 degree rotation around mutual orthogonal axis is a 180 degree rotation.
by iteration of this scheme we can implement
the function 
$
 f_{i;k_0}
$
for every arbitrary $k_0 \in \N_0$. 
Chosing $k_0$ such that 
$
2^{k_0}\geq N
$
we have a measurement procedure for $f_i$.

The number of necessary conditional transformations for 
measuring $f_i$ is  then of the order
$
 2^{k_0}\,,
$
since the measurement of $f_{i;k+1}$ uses
the measurement of $f_{i;k}$ twice.
Hence the complexity of the measurement is linear in $N$, i.e.,
exponential in the number of qubits.

We have already mentioned that measurement procedures
for two functions $f$ and $g$ can concatenated to measurements for
$f\oplus g$. There is also a simple rule to combine two measurements
to obtain $f\wedge g$. For doing so we have to achieve that
$f$ and $g$ implement $u$ and $v$ respectively with the property that
$u$ and $v$  are elements of $S_4$ with the property that
$
w:=u \,v\, u^{-1} \,v^{-1} =u\,v\,u\,v 
$
is an element of the order $2$. 
The group $S_4$ is the symmetry group  of the
cube and the octahedron. 
We can implement the measurement 
for $g$ followed by the measurement of $f$ and repeat both procedures
and we have implemented the transformation
\[
w^{f(j) \wedge g(j)}\,,
\]
 i.e., a measurement for $f\wedge g$. 
Note that this kind of ``computation on a single qubit''
is similar to the so-called ROM-based computation in \cite{TraNiel}.

\section{The simplest measurements}

\label{Simple}

The scheme presented in Section~\ref{Group}  for generic 
functions $f$ involves a rather large number of transformations.
It is straightforward to ask which joint observables are simple to measure.
Referring more specificly to our model it would be interesting to know 
which functions require only rather short sequences of transformations.
The following observation is obvious:
The basic observable in our model is given by $f_{0;1}(S_z)$ 
which was defined as
$f_{0;1}(j):= j\, (mod \,2)$ in Section~\ref{Group}. 
This measurement does not require any
unitary operations on the control spin, we have only to 
initialize the probe spin orthogonal to the $z$ axis and
wait for the time
$t=\pi$. Then the probe spin rotates $180$ degree    
 if the register state is an eigenstate
of  $S_z$ with odd eigenvalue and rotates a multiple of $360^o$
if the eigenvalue is even.
If all spins are coupled to the probe spin with equal interaction strength,
this process measures the parity of the binary word in the register.
If the strengths of the couplings are $1,2,4,\dots$ the process measures
the observable $\sigma_z$ for the spin with the {\it weakest} coupling.
Single qubit-observables for the other qubits seem to be 
less simple to measure.  Note that the other functions
$f_{i;k}$ for $k> 1$ (which are relative elementary in our scheme)
are not single qubit observables
since they refer to the states of at least $k$ qubits.

Now we show how to measure the function $f_{3;2}$ using conditional
and unconditional transformations in the dihedral group $D_8$, the
symmetry group of the octagon.
We have checked that all two-valued functions with 
periodicity $4$ (i.e. $f(j)=f(j+4)$) can 
be measured by a sequence in
\[
D_8\times \cdots \times D_8\,.
\]
Some of the functions with periodicity $4$ 
can be implemented by a sequence of $3$ steps and some functions
can even be measured by sequences of length $2$.
Since it would not give any insights to consider all of them, we
will only describe the procedure for $f_{3;2}$.
For two qubits with coupling strengths $1$ and $2$, respectively, 
this is simply the AND function
(in Section~\ref{Gates} 
we will use this procedure to implement a
controlled-phase-gate on two qubits), 
in general it measures
whether the content of the register is equal to $3$ modulo $4$.
Let $u$ be a rotation by $45$ degree around the $z$ axis. Consider $u$ as
an element of the dihedral group  $D_8$. 
Initialize the probe spin to an arbitrary state in the $(x,y)$-plane.
It is sufficient to consider only the eigenvalues $0,1,2\dots,7$ since 
we use only transformations $w$ with $w^8={\bf 1}$ and therefore 
the probe particle evolves in the same way for $j$ and $j+8$.

Then we use the conditional transformation 
\[
(u^0,u^1,\dots,u^{7})\,.
\]
Fig.~2(a)  
shows the  positions of the resulting states for the eigenvalues 
$j=0,1,\dots,7$.

\begin{figure}
\label{8Pos}
\epsfbox[-15 -15 221 92]{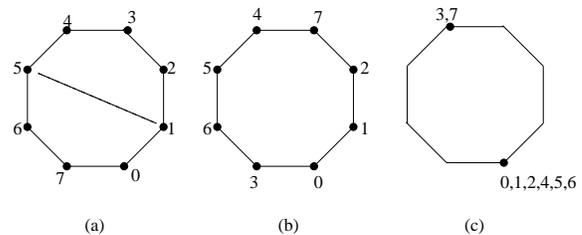}
\caption{Positions of the states for the eigenvalues
$0,1,\dots,7$ after transformations in the dihedral group $D_8$.}    
\end{figure}

Then we implement the conditional transformation 
\[
(v^0,v^1,\dots,v^{7})\,,
\]
where $v$ is a 180 degree rotation in $D_8$ around the axis going through
point 1 and 5  in
Fig.~2(a). The effect is that the state is fixed for all even eigenvalues
and mirrored at the $(1,5)$-line 
for all odd values. The resulting states  are 
shown in Fig.~2(b).

By implementing
\[
(u^{-0},u^{-1},\dots,u^{-7})
\]
the state of the probe spin evolves to the positions shown in Fig~2.(c).
The states for the eigenvalues group in two classes $\{3,7\}$ and 
$\{0,1,2,4,5,6\}$ lying on the opposite side of the Bloch sphere.   
Hence we can distinguish those two classes by measuring the probe spin.
The measurement procedure presented here consists
of $3$ conditional transformations (``steps'').
It would be interesting to know 
which other observables allow simple  measurement procedures.
Note that the only function that can be measured in one step
is the parity function. This is easy to see:
If the conditional transformation
\[
(u^0,u^1,\dots,u^{N-1})
\]
should only contain rotations by 180 degree, one can only have
\[
({\bf 1},u,{\bf 1},u,\dots)\,.
\]

\section{Implementing gates on several qubits}

\label{Gates}

The implementation of gates on $n$ spins by addressing only the control spin
can directly be performed by measuring procedures.
Let $U$ be an $n$-spin
transformation that is diagonal in the $\sigma_z$-basis.
Furthermore assume that $U$ has only two distinct eigenvalues. 
They can assumed to be $e^{-i\phi}, e^{i\phi}$ w.l.o.g. 
 Then we have 
$U=\exp( i \phi f(S_z))$ for some  
function $f: \{0,1,2,\dots,N-1\}\rightarrow \{-1,1\}$.
Apply the following scheme:
(1) Initialize the control spin in the state $|0\rangle$ with respect to the
$\sigma_z$ basis.
(2) Perform a measurement of $f$. 
(3) Implement $\exp(i \phi \sigma_z)$ on the control spin.
(3) Repeat the measurement process (Note that our kind
of measurements are their own inverse).

By selecting interactions $\sigma_x \otimes S_x$ or
$\sigma_y \otimes S_y$ one can similarly obtain unitaries that are
diagonal with respect to the $\sigma_x$ and $\sigma_y$-basis.
For each spin $j$ the single 
qubit transformations 
\[
\exp(i\sigma^{(j)}_\alpha)
\]
with $\alpha=x,y,z$ can clearly be implemented by this method. 
Furthermore the controlled phase gate 
can be implemented using the measurements process
for the AND function of Section~\ref{Simple} if we neglect a physically
irrelevant global phase factor.
With $\phi=\pi $ the controlled phase gate is equivalent to a controlled-not
by single qubit Hadamard transformation.
This shows that
the set of gates that can be obtained is universal for quantum computation.
If one uses functions $f_{i;k}$ with $2^k\geq N$ the process
explained above implements a phase shift controlled by the states
of all qubits.

Note that the results can in principle be generalized to the case that
$S_x,S_y,S_z$ have non-equidistant spectrum. First we find a suitable
rational approximation for the eigenvalues and calculate the 
least common multiple of the denominators. 
By rescaling the  time by this factor
the spectrum can be viewed as a subset of the integers and the algorithm
above can be applied.

\section{Control-spin with different positions}

\label{Flexible}

The measurement and control schemes presented above
are not efficient at all for controlling and measuring
a large number $n$ of qubits since even the computation of simple
functions as the monomials $f_i$ above requires a number of 
operations 
that is linear in $N$ and exponential in $n$ 
in the case that $S_z$ has equidistant eigenvalues.
Now we present a model that is only a slight modification
with respect to its physical assumptions but changes the computational model
significantly: We assume that the controlling electron can be moved
and put at different positions (See Fig.~3). Since the coupling strength
$c_j$ between nucleus  $j$ and electron  depends on the distance $d_j$
between them, the interactions strength is no longer fixed.

\begin{figure}
\epsfbox[-40 0 170 150]{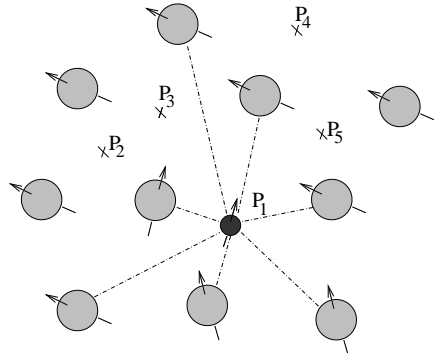}
\caption{The electron can be put at different positions $P_i$.
This changes the coupling to the different nuclear spins.} 
\end{figure}

Note that the
silicon-based  quantum computer
proposed by Kane \cite{Kane}
also uses the spins of mobile electrons 
to control nuclear spins.
However, this analogy should
not be taken to literally since the Kane-proposal 
does not control several nuclei by  the same electron
simultaneously.

For a specific position of the electron, we define the 
``coupling vector''  by
$C:=(c_1,c_2,\dots,c_n)$ where $c_j$ is the strength of the coupling to 
spin $j$. Now we chose $n$ different points $P_1,P_2,\dots,P_n$ 
as possible positions of the electron. In the generic case, we will have that
the corresponding coupling vectors $C_1,\dots,C_n$ are linearly independent.
Then we can achieve an arbitrary effective coupling vector $C$
as follows: If $C=\sum_j d_j C_j$ we
put the electron for the time $t:=|d_j|$ at position $j$ and 
before and after this time interval we implement the transformation
$i\sigma_x$ on the controller spin if and only if $d_j$ is negative.
This reverses the sign of the coupling.

This scheme allows to implement usual two-qubit gates efficiently:
Set for instance 
\[
C:=(0,0,1,0,\dots,0,2,0,\dots,0)
\]
with the values $1$ and $2$  at arbitrary positions $i,j$.
This allows to measure every boolean function of these two qubits
efficiently
or to implement arbitrary two qubit gates.

It may also be reasonable to choose
\[
C:=(0,\dots,0,1,0\dots,0,2,0,\dots,0,4,0,\dots,0) 
\]
in order to control three qubits at once, since the control schemes
above are not too complex for $n=3$.
In order to  compute any function of the Hamming weight of the 
binary word in the  register we choose the effective coupling
$C:=(1,1,\dots,1)$.
Using the results of Sections~\ref{Group} and \ref{Simple} we can therefore 
implement all few qubit operations efficiently.

\section{Comparing the model with the quantum gate model}

To compare the computational power of our model 
to the common model with single and two qubit gates
as elementary operations we have already mentioned in the last Section that
all few qubit gates require complexity $O(n)$ since the electron has
to be put at $n$ different positions in the generic case.

Conversely, the operation
$(d^0,d^1,\dots,d^{N-1})$ which is elementary in our model
(provided that there is a position such that the coupling vector 
is $C=(1,2,2^2,\dots,2^{n-1})$) 
requires $n$ controlled phase shift gates
each acting on one of the nuclear spins and the electron spin. 
Note that we do not consider the
transformations that can be implemented if one uses the
full interaction $H=\sum_\alpha \sigma_\alpha \otimes S_\alpha$.
To investigate the computational power of this interaction may be an 
interesting but
difficult task and the simulation of the corresponding time evolution
by gate operations is non-trivial.

\section{Conclusions}

In principle $n$ spins can be controlled by acting on a 
single control-spin only provided that an appropriate
fixed interaction between controller and the $n$ spins is given.
Measurements and unitary transformations on the
$n$-qubit register can be implemented.
For large $n$, this method becomes rather inefficient.
However, there are still some interesting joint
observables that can be measured efficiently. If all qubits couple with
the probe particle via interactions of the same strength, some
simple functions of the Hamming weight are easy to obtain. 
This illustrates in which way certain joint observables
can relative directly be measured, even more directly than 
single qubit observables.
Our model shows that single qubit observables are not 
necessarily the ``straightforward elementary observables''  
of a complex system. Of course one may object that also our model
relies on single-qubit measurements on the probe particle.
But, if we consider the probe particle as part of the environment
and as more directly observable than the system itself,
we may maintain our statement that other observables
are sometimes more direct than single qubit measurements.
However, the objection  rather shows the
circularity of the deep and interesting
question ``which quantum observables are 
most directly  accessible?''.

\end{multicols}

The authors acknowledge discussions with Martin R\"{o}tteler.
This work is part of the BMBF-project
``Informatische Methoden bei der Steuerung komplexer Quantensysteme''.

\end{document}